\documentclass[%
 graphicx,
 aip,
 jcp,%
 amsmath,amssymb,
 reprint,%
]{revtex4-1}

\usepackage{graphicx}

\begin{document}


\title{Exploring covalently bonded diamondoid particles with valence photoelectron spectroscopy} 



\author{Tobias Zimmermann}
\email[]{tobias.zimmermann@physik.tu-berlin.de}
\author{Robert Richter}
\author{Andre Knecht}
\affiliation{Institut f\"ur Optik und Atomare Physik, Technische Universit\"at Berlin, EW 3-1, Hardenbergstr. 36, 10623 Berlin, Germany}

\author{Andrey A. Fokin}
\affiliation{Department of Organic Chemistry, Kiev Polytechnic Institute, pr. Pobedy 37, 03056 Kiev, Ukraine}
\affiliation{Institute of Organic Chemistry, Justus-Liebig University, Heinrich-Buff-Ring 58, 35392 Giessen, Germany}

\author{Tetyana V. Koso}
\affiliation{Institute of Organic Chemistry, Justus-Liebig University, Heinrich-Buff-Ring 58, 35392 Giessen, Germany}

\author{Lesya V. Chernish}
\author{Pavel A. Gunchenko}
\affiliation{Department of Organic Chemistry, Kiev Polytechnic Institute, pr. Pobedy 37, 03056 Kiev, Ukraine}

\author{Peter R. Schreiner}
\affiliation{Institute of Organic Chemistry, Justus-Liebig University, Heinrich-Buff-Ring 58, 35392 Giessen, Germany}

\author{Thomas M\"oller}
\author{Torbj\"orn Rander}
\affiliation{Institut f\"ur Optik und Atomare Physik, Technische Universit\"at Berlin, EW 3-1, Hardenbergstr. 36, 10623 Berlin, Germany}


\date{\today}

\begin{abstract}
We investigated the valence electronic structure of diamondoid particles in the gas phase, utilizing valence photoelectron spectroscopy. The samples were singly or doubly covalently bonded dimers or trimers of the lower diamondoids. Both the bond type and the combination of bonding partners are shown to affect the overall electronic structure. For singly bonded particles, we observe a small impact of the bond on the electronic structure, whereas for doubly bonded particles, the connecting bond determines the electronic structure of the highest occupied orbitals. In the singly bonded particles a superposition of the bonding partner orbitals determines the overall electronic structure. The experimental findings are supported by density functional theory computations at the M06-2X/cc-pVDZ level of theory.
\end{abstract}

\pacs{33.60.+q, 36.40.Cg, 73.22.-f, 79.60.Jv, 81.07.-b, 82.80.Pv}

\keywords{diamondoids, electronic structure, ionization potential, nanodiamonds, quantum confinement, size-dependence, valence photoelectron spectroscopy}

\maketitle 

\section{Introduction}\label{sec:introduction}

The electronic structure of nanoparticles define their electrical, chemical and optical properties. Hence, investigating these forms the basis for developing new applications in nanotechnology. A comprehensive understanding of the various possibilities to modify the electronic structure of a particle is required when aiming to tailor compounds for specific applications. Research in this area can advance, for instance, the development of electron photoemitters\cite{yang2007monochromatic,ishiwata2012diamondoid} or nano-electronics.\cite{yang2011molecular}

Diamondoids are perfectly size- and shape-selectable, hydrogen passivated, \textit{sp}$^3$-hybridized carbon nanostructures.\cite{schwertfeger2007diamonds} As such, they are an ideal class of particles for the study of effects induced by manipulation of geometry and chemical composition on the electronic structure in nanoscale systems. In addition, functionalization of these particles presents another possibility to tune their electronic structure.\cite{landt2010influence,voros2012tuning,meinke2013experimental,rander2013electronic} Apart from the exploration of their existence in crude oil,\cite{Dahl2003a} continuous improvements in the field of synthesis of diamondoids\cite{schwertfeger2007diamonds} have led to a rise in their popularity during the last ten years. Diamondoids are of interest for the oil industry\cite{Reiser1996160,AliMansoori1997101,dahl1999diamondoid} and environmental protection,\cite{wang2006forensic} and are also used for medicinal applications,\cite{kazimierczuk2001adamantylaminopyrimidines} among others. Of particular interest in nanoparticles is the size-dependence of their electronic structure,\cite{brus1984electron} often referred to in recent literature\cite{PhysRevLett.90.037401,drummond2005electron,PhysRevB.80.161411} as quantum confinement (QC) effects. As they are perfectly size-selectable, diamondoids enjoy great popularity in the investigation of such effects.\cite{willey2005molecular,drummond2005electron,willey2006observation,landt2009optical}

Recently, Schreiner \textit{et al.} synthesized diamondoid particles with extraordinary long CC-bonds.\cite{Schreiner2011,Fokin2012} Except for a study on the smallest\cite{edwards1977anodic} singly bonded particle, their electronic structure have not been investigated. To our knowledge this is the case also for such particles connected with CC-double-bonds,\cite{supplementalmaterial} for which results for only one compound have been reported in the literature.\cite{mollere1976photoelectron} The present work deals with the valence electronic structure of such singly and doubly bonded diamondoid particles. The idea to combine lower diamondoids\cite{mcintosh2004diamond} is comparable to a modular design principle. It is motivated by the question whether such a combination of lower diamondoids can imitate the electronic properties of larger, pristine diamondoids.

\section{Experimental}

We utilized photoelectron spectroscopy (PES) for all measurements in this work. A Scienta SES-2002 hemispherical electron analyzer was used. The samples in the spectrometer focus region were ionized at $21.22$\,eV with a He-lamp (SPECS UVS 10/35). A resistively heated oven was used to bring the samples into the gas phase, see Table~\ref{tab:experimental_parameters} for an overview of the various samples  and the experimental parameters. While the ambient chamber pressure was in the mid $10^{-7}$\,mbar range, chamber pressures during measurements were always kept constant in the mid $10^{-6}$\,mbar range. To calibrate the spectrometer energy scale, and to test the resolution, Xe gas (Air Liquide, 99.995\% purity) was used. The experimental resolution was $50$\,meV for all samples.

\begin{table}
\begin{ruledtabular}
\begin{tabular}{clc}
No.&Sample& Vaporization\\
 & &  Temperature ($^\circ$C)\\
\hline
\textbf{1} & adamantane & $25$\\
\textbf{2} & diamantane & $60-75$\\
\textbf{3} & triamantane & $70-100$\\
\textbf{1--1} & 1-(1-adamantyl)adamantane & $90-100$\\
\textbf{2--2} & 1-(1-diamantyl)diamantane & $100-130$\\
\textbf{3--1} & 2-(1-adamantyl)triamantane & $100-120$\\
\textbf{3--2} & 2-(1-diamantyl)triamantane & $140-160$\\
\textbf{2=1} & adamantylidene-diamantane & $70-110$\\
\textbf{2=2} & diamantylidene-diamantane & $110-140$\\
\textbf{1=2=1} & (di-adamantylidene)-diamantane & $140-170$\\
\textbf{4} & di-pentacycloundecane & $80-110$
\end{tabular}
\end{ruledtabular}
\caption{\label{tab:experimental_parameters} Experimental parameters used for the measurements in this work. Vaporization temperatures are given for ambient chamber pressures in the mid $10^{-7}$\,mbar range.}
\end{table}

\section{Computations}

Computations using density functional theory (DFT)\cite{kohn1965self} were performed for all substances to assist the interpretation of the recorded spectra. We utilized Gaussian09\cite{g09} with the M06-2X functional\cite{zhao2008m06} together with a cc-pVDZ (correlation-consistent) basis set, as this method was previously successfully used to describe the optimized geometries of the singly bonded particles.\cite{Schreiner2011} Second derivatives were computed analytically to confirm that all
structures are minima (NIMAG = 0). All orbital energies and isosurfaces were generated with this method. The stick spectra were convolved with a Gaussian function to take the vibronic broadening and the spectrometer resolution into account. This method shows good agreement with the measured photoelectron spectra for both the pristine diamondoids and the diamondoid dimers, without a rigorous but expensive Franck-Condon analysis for each sample. The adiabatic ionization potentials (IP) were computed by subtracting the total energy of the particle ion from the total energy of the particle ground state. We also observe that the pristine diamondoid IPs are more accurately computed by the M06-2X functional than the B3LYP functional; the latter has been used extensively previously for diamondoid systems.\cite{fokin2009band,landt2010influence,lu2005electronic}

\section{Results}

\begin{figure*}
\includegraphics{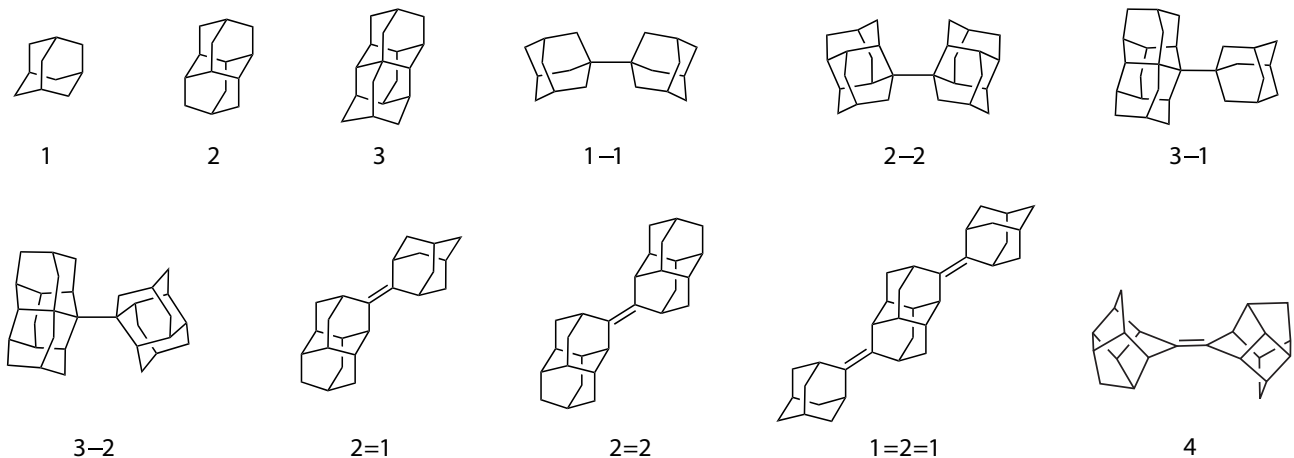}%
\caption{\label{fig:samples}A graphical overview of the samples investigated in this work, as denoted in Table \ref{tab:experimental_parameters}. The lower pristine diamondoids (\textbf{1}-\textbf{3})  are followed by the singly bonded particles (\textbf{1--1}-\textbf{3--2}) and the doubly bonded particles (\textbf{2=1}-\textbf{1=2=1}). Another doubly bonded compound is used for reference purposes (\textbf{4}).}%
\end{figure*}

Measurements were performed on singly and doubly bonded diamondoid particles, see Figure \ref{fig:samples}. Henceforth, we will refer to them by the notation in Table \ref{tab:experimental_parameters}. For the singly bonded particles we investigated homo-dimers, where both diamondoids are of the same type (\textbf{1--1}, \textbf{2--2}), and hetero-dimers, with bonding partners of different types (\textbf{3--1}, \textbf{3--2}).  For doubly bonded particles a hetero-dimer (\textbf{2=1}), a homo-dimer (\textbf{2=2}) and a hetero-trimer (\textbf{1=2=1}) were investigated. For reference purposes, a compound with a CC-double-bond (\textbf{4}) and the first three pristine diamondoids (\textbf{1}-\textbf{3}) were also analyzed. The results will be split in two sections according to the type of bonding.

\subsection{Singly bonded particles}

\begin{figure}
\includegraphics{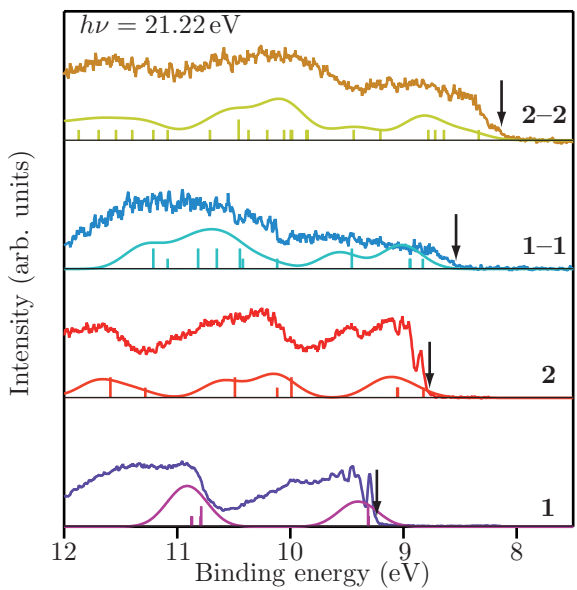}
\caption{\label{fig:graph1}The valence spectra of the two singly bonded homo-dimers (top) and their corresponding pristine monomers (bottom). The ionization potentials, denoted by black arrows in the figure, change to lower binding energies for the diamondoid particles towards the single molecules.}
\end{figure}

The valence photoelectron spectra of \textbf{1}, \textbf{2}, \textbf{1--1} and \textbf{2--2} are shown in Figure \ref{fig:graph1}, together with the computed spectrum of each sample. Comparing the spectra of the dimers with the respective monomers reveals that the overall structures of the photoelectron bands show similarities in both cases, e.g., the number of individual bands, although the dimers have broader bands shifted to lower binding energies and lack the distinct ionization onset present in the monomers.

All IPs presented in this work are adiabatic ionization potentials and were determined by the method described by Lenzke \textit{et al.}.\cite{Lenzke2007} For pristine diamondoids the change in IP with the diamondoid size is well known and has been ascribed to QC effects.\cite{Lenzke2007} The diamondoid particles show a more complex geometry compared to the pristine diamondoids. However, the diamondoid particle IPs also show a decrease with increasing size, as can be seen in Figure \ref{fig:graph1}. Hence, diamondoid dimers seem to be subjected to QC effects as well.

\begin{figure}
\includegraphics{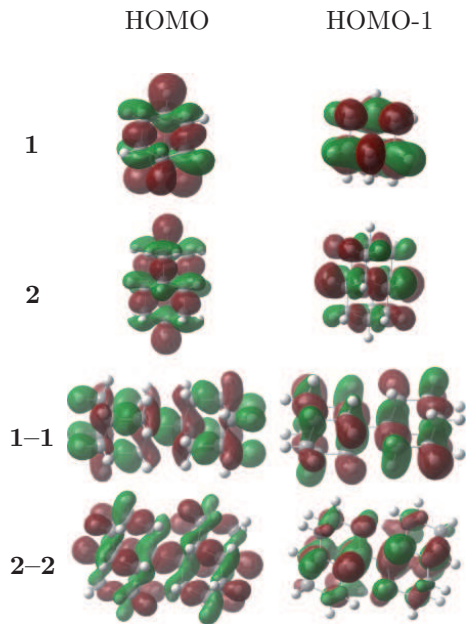}
\caption{\label{fig:orbitals_homodimers}The HOMOs (left) of the two singly bonded homo-dimers (bottom) and their corresponding pristine monomers (top) in comparison to their HOMO-1 (right). The M06-2X functional together with a cc-pVDZ basis set was utilized for the computations. An isovalue of $0.02$ is used for all isosurfaces presented in this work.}
\end{figure}

Figure \ref{fig:orbitals_homodimers} shows the computed highest occupied molecular orbitals (HOMOs) of the singly bonded diamondoid homo-dimers together with their associated monomers. The orbitals are delocalized and symmetrically distributed over the entire cage structures, i.e., there is no distinction between individual CC-bonds in the system, for both monomers and dimers. Calculations show that this is also true for the deeper lying valence orbitals in the case of equal bonding partners. The dimerization of two identical pristine diamondoids leads to an increase of the orbital volume in comparison to the situation of the monomer by a factor of roughly two. This resembles the situation in pristine diamondoids where the HOMO volume also increases with size. This volume increase leads to a lowering of the IP, and can be ascribed to QC.\cite{willey2006observation} Optimized geometries for the singly bonded dimer ions show a considerable elongation of the central bond of up to $1$\,\r{A}, indicating dissociation upon ionization. This explains the flat and featureless ionization onsets of the homo-dimers in comparison to the monomers.

\begin{figure}
\includegraphics{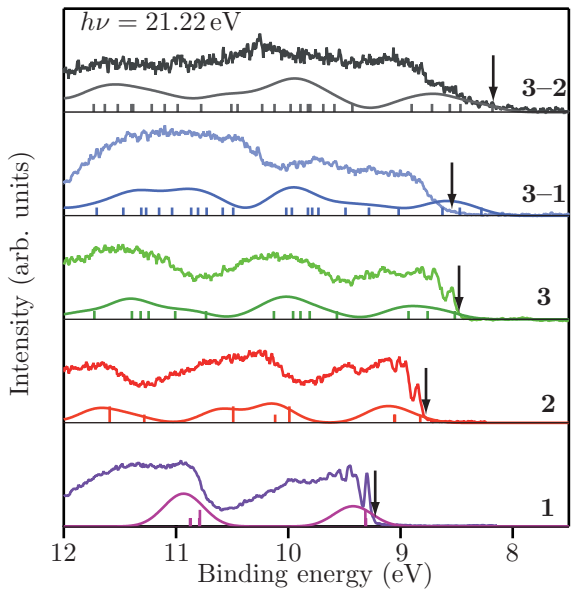}
\caption{\label{fig:graph2}The valence spectra of the two singly bonded hetero-dimers (top) and their corresponding pristine monomers (bottom). In contrast to the homo-dimers the ionization potentials, denoted by black arrows in the figure, do not strictly shift to lower energies but depend on the bonding partners involved.}
\end{figure}

\begin{figure}
\includegraphics{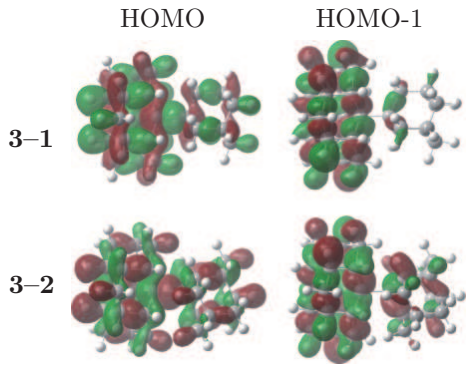}
\caption{\label{fig:orbitals_heterodimers}The HOMOs (left) of the two singly bonded hetero-dimers \textbf{3--1} and \textbf{3--2} in comparison to their HOMO-1 (right). The M06-2X functional together with a cc-pVDZ basis set was utilized for the computations.}
\end{figure}

Studying the change of IPs for diamondoid hetero-dimers (Figure \ref{fig:graph2}), we observe a distinct behavior for dimers containing different bonding partners. Both studied hetero-dimers include triamantane (\textbf{3}) as one bonding partner. In combination with adamantane (\textbf{1}) the dimer (\textbf{3--1}) IP is significantly lower than for dimer \textbf{3--2} containing diamantane (\textbf{2}). However in both cases the dimer IPs differ only slightly from the IP of pristine triamantane. The computed HOMOs for the two hetero-dimers (Figure \ref{fig:orbitals_heterodimers}) still show delocalization over the whole dimer, but with a tendency towards localization on the triamantane moiety. For \textbf{3--1}, the HOMO-1 is nearly completely localized to the triamantane part of the dimer. In the case of \textbf{3--2} this localization is less pronounced and the orbital extends also to the smaller bonding partner side. A comparison of the IPs for the bonding partners in their pristine form (Table \ref{tab:ip_single}) reveals that the HOMO of \textbf{1} differs by about $0.75$\,eV from \textbf{3} while the difference is only $0.3$\,eV for the IPs of \textbf{2} and \textbf{3}. This difference in binding energies of the monomer highest occupied orbitals leads to a higher localization in the triamantane moiety than in the adamantyl or diamantyl moieties of the dimers (Figure \ref{fig:orbitals_heterodimers}). The confinement of the orbital is higher for \textbf{3--1} than for \textbf{3--2} and therefore the IP of \textbf{3--2} is lower. The computations thus suggest that the hetero-dimer electronic structures are a superposition of the monomer orbitals involved. With exception of \textbf{3--1}, the agreement between the measured data and computations is good. This indicates that in most cases an approach of combined orbitals is applicable to singly bonded diamondoid dimers, regardless of whether the bonding partners are of the same type or not.

\begin{table*}
\begin{ruledtabular}
\begin{tabular}{clcc}
No.&Compound&Experimental IP (eV)&Computed IP (eV)\\
\hline
\textbf{1} & adamantane & $9.24(5)$ & $9.32$\\
\textbf{2} & diamantane & $8.79(5)$ & $8.89$\\
\textbf{3} & triamantane & $8.49(5)$ & $8.58$\\
\hline
\textbf{1--1} & 1-(1-adamantyl)\-adamantane & $8.59(5)$ & $8.41$\\
\textbf{2--2} & 1-(1-diamantyl)\-diamantane & $8.19(5)$ & $7.82$\\
\hline
\textbf{3--1} & 2-(1-adamantyl)\-triamantane & $8.65(5)$ & $7.71$\\
\textbf{3--2} & 2-(1-diamantyl)\-triamantane & $8.27(5)$ & $7.27$
\end{tabular}
\end{ruledtabular}
\caption{\label{tab:ip_single}Measured and M06-2X/cc-pVDZ computed IPs for singly bonded diamondoid particles and respective pristine diamondoids.}
\end{table*}

Table \ref{tab:ip_single} summarizes the measured IPs together with the computed IPs of the singly bonded particles presented in this work. The agreement between the experimental and the computed values is almost within range of the experimental error for the pristine diamondoids. Even though for the dimers, the experimental and computed values differ up to roughly 1 eV, the computations reflect the decrease of IP with increasing bonding partner size accurately. Further study using more elaborate computations would be needed to gain better absolute agreement with experimental values.

\subsection{Doubly bonded particles}

\begin{figure}
\includegraphics{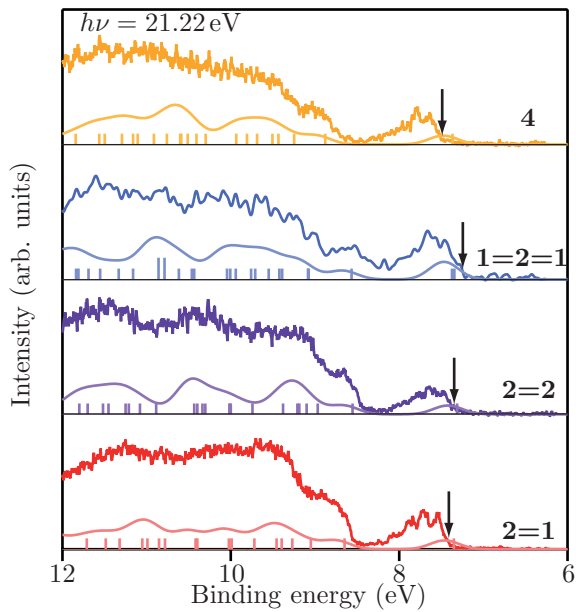}
\caption{\label{fig:graph3}The valence spectra of the doubly bonded substances. The spectrum of \textbf{1=2=1} was smoothed for presentation. The IPs, denoted by black arrows, vary only slightly for the doubly bonded particles presented in this work.}
\end{figure}

The photoelectron spectra of the doubly bonded particles, shown in Figure \ref{fig:graph3}, differ mainly in the ionization onset region from the singly bonded particles; here, an isolated $\pi$-band with vibrational fine structure can be seen. This vibrational progression can be ascribed to the C=C-stretch mode, well known from the photoelectron spectra of other alkenes.\cite{mintz1979photoelectron} The fact that the relative intensity of the first vibrational band is higher in the spectrum of \textbf{1=2=1} than for the other compounds supports the assignment of this feature to the double bond as there are two of them in that case. Studies on the photoluminescence of hydrogenated amorphous carbon show that the $\pi$-states of \textit{sp}$^2$-sites in an overall \textit{sp}$^3$-matrix form the valence-band edge.\cite{robertson1996photoluminescence,robertson1996recombination} The dominant ionization onset of the \textit{sp}$^2$-feature in the doubly bonded diamondoid particles indicate similar behavior in a few-atom size regime, far away from the bulk. The energetic position from this band changes only slightly with particle size, and therefore, the IPs of the doubly bonded structures lie within a small energy region around $7.2$\,eV (Table \ref{tab:ip_double}). At higher binding energies, overlapping bands conceal any distinctive features. Only at the edge of these bands, a shoulder is visible for each substance. A similar shoulder is also known for other alkenes and it has been assigned to the $2p\sigma$-electrons for 2,3-dimethylbutene ("tetramethylethylene"),\cite{mintz1979photoelectron} whose structure is comparable to the central CC-double-bond and the next four neighbor atoms.

\begin{figure}
\includegraphics{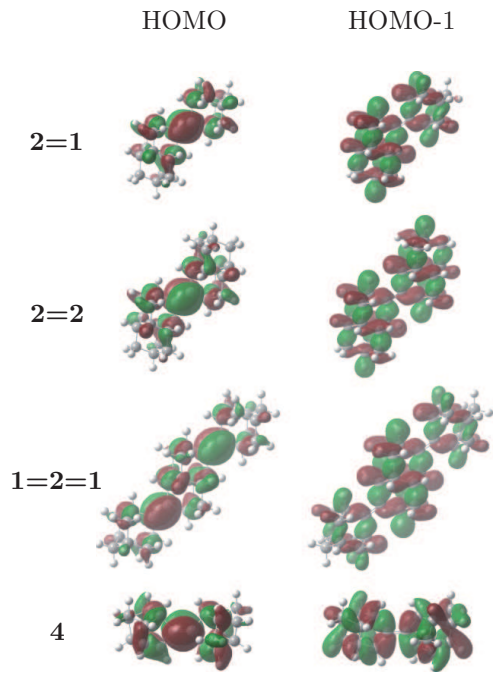}
\caption{\label{fig:orbitals_double}The HOMOs (left) of the doubly bonded diamondoid particles (\textbf{2=1}-\textbf{1=2=1}) and a doubly bonded reference compound (\textbf{4}) and the HOMO-1 (right). The M06-2X functional together with a cc-pVDZ basis set was utilized for the computations.}
\end{figure}

Computations of the HOMOs (Figure \ref{fig:orbitals_double}) show a strong localization to the CC-double-bonds. The probability density at the diamondoid cages is low for the HOMO which explains the weak dependence on the overall size of the particles. QC effects are reduced due to the fact that the $\pi$-electrons are nearly unaffected by the sizes of the surrounding diamondoid cages. In contrast, the HOMO-1 is distributed over the $\sigma$-bonds of the diamondoid cages and therefore a size dependence of this orbital energy can be expected. The small differences of the measured IPs for the particles under consideration can be explained with the screening of the $\pi$-electrons by the surrounding $\sigma$-electrons. This results in a relatively constant HOMO energy in comparison to the shift of the HOMO-1 energy. Hence, the energetic distance of the HOMO and HOMO-1 decreases with increasing size of the bonding partners. From studies on pristine diamondoids a decrease of QC effects with increasing diamondoid size is known.\cite{Lenzke2007} Thus, for larger bonding partners, a saturation of the decreasing energetic distance of HOMO and HOMO-1 is expected at some point. Further study may show if this saturation  leads to a change of energetic ordering of the uppermost occupied orbitals.

The analysis of \textbf{4} indicates that these characteristics apply not only for doubly bonded diamondoid particles but also for \textit{sp}$^3$-hybridized molecules joined by CC-double-bonds in general. With 22 carbon atoms being the smallest doubly bonded substance of the study, compound \textbf{4} shows the highest IP of the analyzed samples.

\begin{table*}
\begin{ruledtabular}
\begin{tabular}{clcc}
No.&Compound&Experimental IP (eV)&Computed IP (eV)\\
\hline
\textbf{2=1} & adamantylidene-diamantane & $7.43(5)$ & $7.35$\\
\textbf{2=2} & diamantylidene-diamantane & $7.32(5)$ & $7.29$\\
\textbf{1=2=1} & (di-adamantylidene)-diamantane & $7.22(5)$ & $7.56$\\
\textbf{4} & di-pentacycloundecane & $7.47(5)$ & $7.47$
\end{tabular}
\end{ruledtabular}
\caption{\label{tab:ip_double}Measured and M06-2X/cc-pVDZ computed IPs for doubly bonded diamondoid particles and a reference compound.}
\end{table*}

\section{Conclusions}

We have studied the valence electronic structure of diamondoid particles. The influence of the bonding partners and the types of connecting bond have been investigated. For singly bonded particles, the central CC-bond has only little impact on the energy levels of the dimers under consideration. Moreover, we observe that a combination of the bonding partner orbitals describes the overall electronic structure well. A consequence of this combination process can be seen through an analysis of the particle IPs. While for the homo-dimers we measure IPs well below the corresponding monomers, the change of IPs for hetero-dimers strongly depends on the particle composition. QC effects can be seen for the homo-dimer IPs but seem to be nearly absent for IPs of hetero-dimers. DFT computations show symmetrically distributed valence orbitals for dimers with equal bonding partners. As in pristine diamondoids, the HOMOs are delocalized over the entire molecule. In hetero-dimers, the constituent orbitals are asymmetrically distributed. The HOMO tends to be localized to the larger bonding partner due to its lower IP. Hence the overall IP resembles that of the larger bonding partner, and QC effects are less pronounced.

For doubly bonded particles, the CC-double-bond has a clear and identifiable impact on the electronic structure. It appears as an isolated electronic feature with vibrational fine-structure at the ionization onset in the photoelectron spectra. The IPs vary only slightly with the size of the constituents. Computations show a strong localization of the HOMO to the CC-double-bond and the influence on the HOMO of the surrounding carbon atoms is restricted to screening. The HOMO-1 is delocalized over the $\sigma$-bonds of the diamondoid cage structures and can be identified as a distinct shoulder in the photoelectron spectra.

To further understand the influence of the combination of single diamondoids to larger particles on the electronic structure of the resulting system, more studies on this class of compounds are highly desirable. Using singly bonded particles enables researchers to imitate the electronic properties of higher diamondoids, thereby circumventing the problem of extremely low synthesis/isolation yields for higher diamondoids. Furthermore, the study of singly bonded dimers is an interesting route to gain insight into the role of dispersive forces on the long central CC-bonds. The further investigation of doubly bonded particles assists the comprehension of carbon nanostructures with both \textit{sp}$^3$- and \textit{sp}$^2$-hybridized moieties. Besides the already explored parameters size, shape, and functionalization, the combination of pristine diamondoids into particles with different bonding situations opens a completely new degree of freedom to explore with regards to electronic structure and band gap tuning.


%
%

%

\begin{acknowledgments}
We would like to thank the Deutsche Forschungsgemeinschaft DFG for financially supporting this work through grant FOR 1282 MO719/10-1 and the Ministry of Science and Education of Ukraine for support through grant 2520f. We would also like to thank Professor Peter Zimmermann for fruitful discussion and assistance with the experimental setup.
\end{acknowledgments}

%

\end{document}